\documentclass[12pt]{article}

\usepackage{PRIMEarxiv}
\usepackage[table,usenames,dvipsnames]{xcolor}\usepackage[utf8]{inputenc} 
\usepackage[T1]{fontenc}    
\usepackage[colorlinks=true,pagebackref=true,linkcolor=BrickRed,citecolor=OliveGreen]{hyperref}       
\usepackage{url}            
\usepackage{booktabs}       
\usepackage{amsfonts}       
\usepackage{nicefrac}       
\usepackage{microtype}      
\usepackage{lipsum}
\usepackage{fancyhdr}       
\usepackage{graphicx}       
\graphicspath{{media/}}     

\usepackage[small,bf]{caption}
\usepackage{float}
\usepackage{subcaption}
\usepackage{tabularx}
\usepackage{multirow}
\usepackage{amsmath}
\usepackage{fontsize}
\usepackage{parskip}
\usepackage{booktabs}
\newcolumntype{R}{>{\raggedleft\arraybackslash}X}
\newcolumntype{L}{>{\raggedright\arraybackslash}X}
\newcolumntype{Y}{>{\centering\arraybackslash}X}
  
\title{Visual attention information can be traced on cortical response but not on the retina: evidence from electrophysiological mouse data using natural images as stimuli
}

\author{
  Nikos Melanitis, Konstantina Nikita \\
  BIOSIM laboratory\\
  National Technical University of Athens \\
  \texttt{email@email} \\
}

\changefontsize{12}
\begin{document}
\maketitle
\begin{abstract}

Visual attention forms the basis of understanding the visual world. In this work we follow a computational approach to investigate the biological basis of visual attention. We analyze retinal and cortical electrophysiological data from mouse. Visual Stimuli are Natural Images depicting real world scenes. Our results show that in primary visual cortex (V1), a subset of around $10\%$ of the neurons responds differently to salient versus non-salient visual regions. Visual attention information was not traced in retinal response. It appears that the retina remains naive concerning visual attention; cortical response gets modulated to interpret visual attention information. Experimental animal studies may be designed to further explore the biological basis of visual attention we traced in this study. In applied and translational science, our study contributes to the design of improved visual prostheses systems- systems that create artificial visual percepts to visually impaired individuals by electronic implants placed on either the retina or the cortex.
\end{abstract}


\section{Introduction}

Visual attention, attracted to the most salient regions in visual space, lies in the core of understanding the visual world. Computational saliency maps have long been introduced~\cite{itti1998model} and since repeatedly improved~\cite{cornia2018predicting, gupta2020salient, borji2018saliency}. In terms of computational sciences, saliency maps are an intermediate representation towards core computer vision tasks~\cite{li2015visual}: semantic segmentation~\cite{gan2015devnet}, event detection~\cite{shimoda2016distinct}, image cropping~\cite{rother2006autocollage} and summarization~\cite{simakov2008summarizing} as well as visual recognition tasks~\cite{rutishauser2004bottom} and image classification~\cite{wu2013scale}. In visual prosthesis~\cite{weiland2014retinal, fernandez2018development}-an intervention in which an implanted device delivers electrical stimuli to compensate the lost sense of vision and evoke visual representations to the implantee- predicting the loci of visual attention-ie eye fixations- may be exploited to improve the prosthetic vision~\cite{alevizaki2019predicting, melanitis2019biologically}. 

Evidence about the ‘where’ of biological visual attention~\cite{treue2003visual, moore2015and} has been collected in numerous studies. Diverging hypotheses have been formed on the location of a biological saliency map: some studies point to the parietal cortex~\cite{gottlieb1998representation} or the thalamus\cite{robinson1992pulvinar,treue2003visual}, however more evidence points to the primary visual cortex V1~\cite{li2002saliency,  treue2003visual, treue2001neural, treue1999feature}. Computational analysis  of retinal and cortical response at single-neuron precision to complex stimuli such as natural images would provide more support on the distribution of the biological saliency map and the characteristics of attention influences.

In this work we are looking for attention influences on two visual prosthesis implantation sites: retina~\cite{da2016five, benetatos2021assessing} and V1~\cite{fernandez2021visual, pouratian2019early}. We analyze the response of biological neurons to natural images from Imagenet~\cite{deng2009imagenet}: retinal ganglion cells (RGC) response is generated from a deep model trained on RGC responses to natural images~\cite{lozano20183d, papadopoulos2021machine}, cortical response is measured experimentally~\cite{stringer2019high}. Using statistical tools we find that a subpopulation of V1 neurons respond differently to salient and non-salient image regions.

\section{Methods}
\subsection{Datasets}

\subsubsection{V1 response Data (Mouseland)}
We use Mouseland~\cite{stringer2019high}, a public dataset of simultaneously recorded V1 cell responses from mouse cortex. The same set of $2800$ natural images was shown to four living mice. Each image was shown at least two times in each recording session.  We use seven recording sessions (Table~\ref{tab:V1sessions}). Natural Images were obtained from Imagenet dataset~\cite{deng2009imagenet}- a cornerstone dataset introduced for object-categorization. Resonance-scanning two-photon calcium microscopy allowed to simultaneous record V1 cells. 
\begingroup
\hyphenpenalty 10000
\exhyphenpenalty 10000
\begin{table}
\footnotesize
\begin{center}
\begin{tabularx}{0.99\linewidth}{L  L L L L L L L L}
\midrule[1.5pt]
Session & $1$ & $2$ & $3$ & $4$ & $5$ &$6$ & $7$  \\ \midrule[1.5pt] 
 Number of recorded cells &$11449$ &$14062$& $9410$& $8122$ & $8704$ &$10145$ &$10103$ \\
\lasthline
\end{tabularx}
\caption{Total number of V1 neurons simultaneously recorded in each of the seven sessions analyzed from Mouseland Dataset.}
\label{tab:V1sessions}
\end{center}
\end{table}
\endgroup

\subsubsection{Retina response Data}
The set contains recordings from $60$ biological (mouse)
RGCs. Recordings were made on the retina recovered from a euthanized mouse using an electrode array~\cite{lozano20183d}. Natural Images of $50 \times 50$ pixels are projected
on the retina for $50$ms each and for a total duration of
$4$min. The sequence starts with $300$ms of darkness. The
set contains static images only. We record cell response
every $10$ms. In total, we get $24479$ recordings per cell. 

We get twelve reliably
recorded RGCs~\cite{papadopoulos2021machine, melanitis2021using}, through Spike Triggered Average (STA)
analysis~\cite{chichilnisky2001simple}. Errors in raw data processing (e.g. in spike
sorting) and/or at the retina preparation may corrupt
the biological recordings. We trained a Convolutional
Neural Network (CNN) model~\cite{mcintosh2016deep} on the set and
then fed white noise sequences to the model to get an
unbiased Receptive Field (RF) estimate through STA~\cite{chichilnisky2001simple}.
Reliable cells were selected based on spatial (center surround
antagonism) and temporal (biphasic response)
STA characteristics. The STA properties of RGCs have
been documented in the literature~\cite{chichilnisky2001simple}.

\subsubsection{Retina Saliency Dataset}
In our analysis of visual attention and retinal response we use Toronto Dataset~\cite{bruce2007attention}, a set of $120$ indoor and outdoor images listed in numerous saliency benchmarks\footnote{MIT saliency benchmark \url{http://saliency.mit.edu/datasets.html}}. We get RGCs spiking response on this set by our retina model and generate the saliency maps from the Saliency Attentive Model (SAM)  (see below, Section~\ref{sec:sam}).
\subsection{Receptive Field (RF) estimation}
\begin{figure}[t]
\setlength\partopsep{-\topsep}
    \centering
    \includegraphics[height=0.25\textheight]{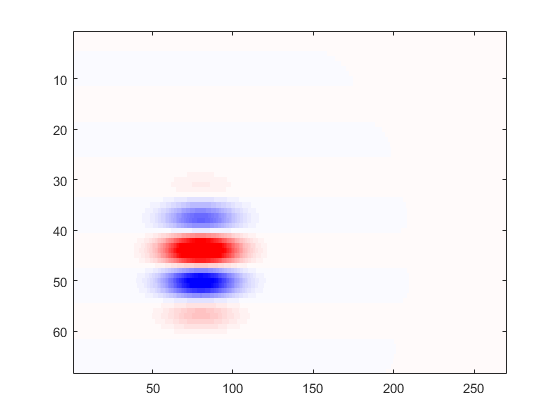}
    \caption{Demonstration of Gabor Receptive Field estimation in Mouseland dataset for a V1 neuron.
}
    \label{fig:GaborRF}
\end{figure}

We evaluate RF shape and location in the visual field for each V1 cell~\cite{touryan2005spatial}. We use a Gabor-filter model to simulate cell response to natural images; we search for an optimal Gabor filter optimizing the explained variance of the model~\cite{smyth2003receptive} (Figure~\ref{fig:GaborRF}).

To get RGCs RFs we trained a Convolutional Neural Network (CNN) model~\cite{mcintosh2016deep}  and then fed white noise sequences to the model to get an unbiased RF estimate through STA~\cite{chichilnisky2001simple, melanitis2021using}.

\subsection{Saliency Estimation}
\label{sec:sam}
\begin{figure}[t]
\setlength\partopsep{-\topsep}
     \centering
     \begin{subfigure}[b]{0.5\textwidth}
         \centering
         \includegraphics[width=\textwidth]{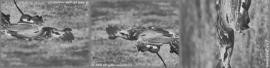}
     \end{subfigure}
     \\ \vspace{5pt}
     \begin{subfigure}[b]{0.5\textwidth}
         \centering
         \includegraphics[width=\textwidth]{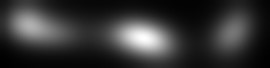}
     \end{subfigure}
        \caption{ Example of an Input Image from Mouseland Set (\textbf{top}) and the SAM estimated saliency map (\textbf{bottom}). In Mouseland, each image taken from Imagenet set is projected on three screens surrounding the Mouse’s head~\cite{stringer2019high}. }
        \label{fig:Saliency}
\end{figure}

We estimate visual saliency using Saliency Attentive Model (SAM)~\cite{cornia2018predicting}, a Deep Learning (DL) model based on Long Short Term Memory (LSTM) recurrent neural networks. Recurrent networks iteratively refine saliency estimations. Multiple learned priors make SAM model able to learn the bias in eye fixations around the image center. SAM model is optimized using a sophisticated multi-term loss function that has been specially designed to  reflect visual attention properties~\cite{cornia2018predicting}. In this paper the SAM model extracts image features using a pre-trained ResNet50 network~\cite{he2016deep} trained on SALICON image dataset~\cite{huang2015salicon} (Figure~\ref{fig:Saliency})).

\subsection{Joint analysis of cortical (V1) response and visual attention}
We examine the effect of visual saliency on neuronal firing rate using Kolmogorov-Smirnov (KS) statistical test~\cite{massey1951kolmogorov, press2007numerical}. KS test is a nonparametric method to see if two arbitrary distributions are the same, based on comparing cumulative distribution functions (CDFs). More specifically, let:
\begin{equation}
F^1, F^2
\end{equation}
be two CDFs and let $F_n$ be the empirical distribution function
\begin{equation}
F_n(t) = \frac{1}{n} \text{number of elements in the sample $\leq t$}
\end{equation}
of $n$ observed data points. To compare $F^1,F^2$ the KS test computes the statistic:
\begin{equation}
D_n(x) = \max_x |F_n^1(x)-F_n^2(x)|
\end{equation}
If two distributions $F^1,F^2$ are equal, then $D_n$ would approach $0$, if the distributions do not overlap at all $D_n$ will approach its maximum value of $1$.

In this work we compare two datasets to see if they are significantly different, so the empirical distribution functions in KS test are chosen as:
\begin{enumerate}
\item $F^1$: firing rate of neuron when the neuron observes maximum saliency image region
\item $F^2$: firing rate of neuron when the neuron does not observe maximum saliency image region
\end{enumerate} 
We conduct the aforementioned KS test for each V1 neuron in the dataset. Following this procedure, we can tell for each neuron whether the neuronal responses at salient image regions are statistically distributed differently from the neuronal responses at non-salient image regions.

\begin{figure}[t]
\setlength\partopsep{-\topsep}
    \centering
    \includegraphics[height=0.15\textheight]{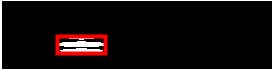}
    \caption{Selecting a Rectangle (red outline) enclosing the Receptive Field. We show a V1 neuron’s Gabor-estimated Receptive Field (white) overlaid on the Mouseland Image space (black).
}
    \label{fig:GaborRec}
\end{figure}

We decide whether a V1 neuron observes the maximum saliency image region applying the following procedure: (i) we take the smallest rectangle that encloses the neuron’s Gabor RF (Figure~\ref{fig:GaborRec}), (ii) we tile the image’s saliency map with the rectangle RF, (iii) we get the average saliency at each tile and finally (iv) we check if the neuron is located on the maximum saliency tile.

In Mouseland dataset each image is shown twice; we get the average Firing Rate (FR) over these two presentations.

We examine correlation between saliency and neuronal response by taking Pearson ($r$) and Spearman ($\rho$) correlation coefficients and  Kendall tau ($\tau$)~\cite{press2007numerical}.  Similarly to KS test, we correlate the neuron's average FR over the two image presentations with the average saliency in the smallest rectangle image region enclosing the neuron’s Gabor RF.

\subsection{Joint analysis of retinal response and visual attention}
\subsubsection{Optimized Firing Rate generation from retinal model}
\begin{figure}[t]
\setlength\partopsep{-\topsep}
     \centering
     \begin{subfigure}[b]{0.5\textwidth}
         \centering
         \includegraphics[width=\textwidth]{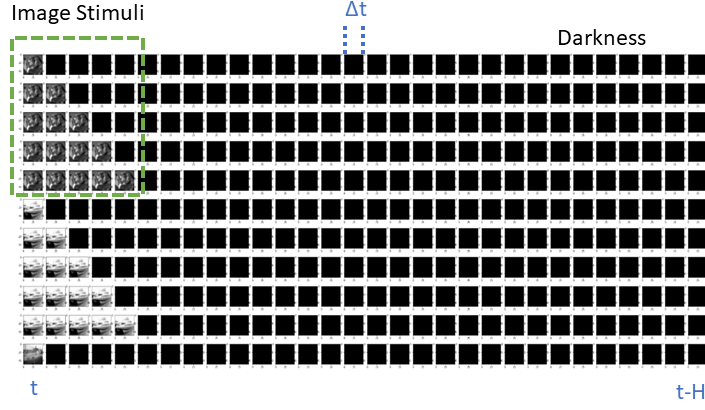}
     \end{subfigure}
     \\ \vspace{5pt}
     \begin{subfigure}[b]{0.5\textwidth}
         \centering
         \includegraphics[width=\textwidth]{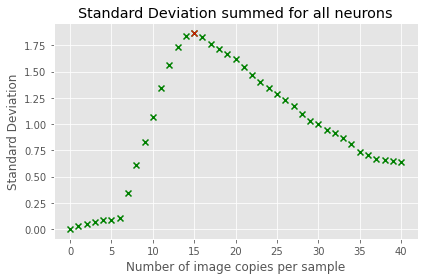}
     \end{subfigure}
        \caption{ Presenting Natural Images to Retina CNN model. \textbf{Top:} The CNN model predicts the firing rate at time bins (each bin lasting $\Delta t = 10$ms). At a given moment, CNN response is a function of stimuli presented over a time period $H$ ($H=400ms$). We present each Image for a $t^*$ time period. Every row in the figure represents an input to the CNN model. We repeatedly present stimuli for a time $t^*$. Image stimuli are preceded by darkness. (Adjusted from~\cite{papadopoulos2021machine}). \textbf{Bottom:} We choose $ t^*$ so that the variance of the response is maximized ($t^*=150$ ms).  We sum the response standard deviation for all twelve RGCs we modeled (on the vertical axis). Each Image copy per sample (horizontal axis) is shown to the model for $10$ms. Standard deviation sum is maximized at $15$ copies, corresponding to $t^*=150$ms. }
        \label{fig:CNNtime}
\end{figure}

The CNN retina model generates a timecourse of RGC FRs, which we need to reduce to a scalar FR value to use in our analyses. We are looking for a timepoint $t^*$, counting from the presentation of an image  ($t=0$), to sample RGC response at $t^*$. We analyze responses of all twelve RGCs and set $t^*$ so that the variance of RGC responses is maximized. RGC response to stimuli is biphasic~\cite{sousa2009bioelectronic}. So RGC response varies little over different inputs at $t$ that fall in the flat areas of RGC response curve. In contrast, in the neighborhood of response curve peaks, FR varies greatly between different inputs. We find that $t^*=150$ms, which corresponds to showing the input image for $150$ms before measuring the retinal response (Figure~\ref{fig:CNNtime}).
\subsubsection{Analysis tools: Firing Rate ratios and correlation coefficients}
To analyze the effect of visual attention on retinal response we use $r,\rho,\tau$ correlation coefficients, as discussed in the previous sections, but also two FR ratios we introduce here. Our objective is to examine if FR is modulated by saliency, so we introduce the following ratios which compare FR at the most salient image region (see Image Splitting, Section~\ref{sec:is}) to a baseline FR. The ratios are evaluated for an input image $i$ and an output RGC $j$ as:
\begin{align}
r_{img} &= \frac{\text{FR of RGC $j$ in most salient  patch in image $i$}}{\text{Average FR of RGC $j$ over all patches in image $i$}} \\
\nonumber \\
r_{rgc}&= \frac{\text{FR of RGC $j$ in most salient  patch in image $i$}}{\text{Average FR of RGC $j$ over all  images}}
\end{align}
In both ratios we compare the RGC response in the most salient image region with the average response: i) of the RGC across all image regions ($r_{img}$) or ii) of the RGC across all image regions in all images ($r_{rgc}$).
\subsubsection{Image Splitting}
\label{sec:is}
\begin{figure}[t]
\setlength\partopsep{-\topsep}
     \centering
     \begin{subfigure}[b]{0.45\textwidth}
         \centering
         \includegraphics[width=\textwidth]{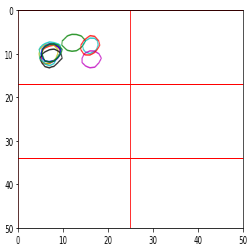}
     \end{subfigure}
     ~
     \begin{subfigure}[b]{0.45\textwidth}
         \centering
         \includegraphics[width=\textwidth]{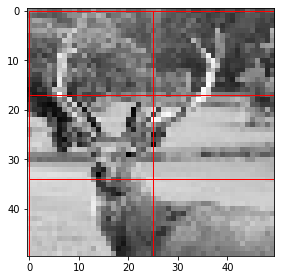}
     \end{subfigure}
        \caption{We show the Receptive Fields (RFs) of reliably-recorded RGCs in $50\times50$ pixels Images. RGCs are contained in a $17\times25$ pixels Image subregion (\textbf{left}). Each input image is split in  six ($3\times2$ grid) rectangle patches (\textbf{right}).}
        \label{fig:RetRFs}
\end{figure}

Each Image is split in six ($3\times2$ grid) rectangle patches (Figure~\ref{fig:RetRFs}). In the original images, all twelve RGC RFs are concentrated in the upper left image corner. So we take the smallest image rectangle which encloses all twelve RFs and tile the image, resulting in six patches per image. Each of the six image patches is separately fed to the retina model to yield responses of RGCs situated within each patch, resulting in six responses for each of the twelve RGCs in each image.
\section{Results}
\subsection{In V1, a subset of cells responds differently to salient stimuli}
We analyzed Mouseland neuronal Data on ImageNet with SAM-produced Saliency maps. In Table~\ref{tab:V1sessions} we give the number of V1 neurons simultaneously recorded in each of the seven experimental sessions. We do a KS test for each neuron in each session comparing the distribution of responses in salient and non-salient image regions. In Table~\ref{tab:V1ks} we gather the fraction of neurons in each session for which the null hypothesis
\begin{equation}
Ho:\text{data come from populations with the same distribution}
\end{equation}
is rejected, ie neuron responds differently to salient and unsalient image regions. Significance level is $p=0.05$. $p$-values are considered accurate if~\cite{press2007numerical}:
\begin{equation}
\label{eq:pacc}
\frac{n_1n_2}{n_1+n_2}\geq 4
\end{equation}
where $n_1,n_2$ are the number of samples in the two sets in KS test. In Table~\ref{tab:V1ksaccu} we give the updated~\ref{tab:V1ks} Results when we enforce condition~\ref{eq:pacc} to ensure p-value accuracy. 

A requirement of the KS test conducted is that a neuron  observes the most salient image region at least once in the image set, otherwise the  method will fail. This happens for a varying number of neurons shown in Table~\ref{tab:V1ksfcase}. We correct the total number of neurons in each session by removing fail cases and refine the results in Tables~\ref{tab:V1ksexamined}, \ref{tab:V1ksexaminedaccu}. 

We observe that a subset of V1 neurons, ranging from $10\%$ to $15\%$ of the cells, responds differently to salient and non-salient image regions.

In Table~\ref{tab:V1cc} we give the range of values of $r,\rho,\tau$ correlation coefficients. We see that correlation analysis is not informative and does not reveal a relationship between saliency and neuronal response.

\begingroup
\hyphenpenalty 10000
\exhyphenpenalty 10000
\begin{table}
\footnotesize
\begin{center}
\begin{tabularx}{0.99\linewidth}{L  L L L L L L L}
\midrule[1.5pt]
$H_0$ rejected &	$0.1039$ &   $0.0984$ &  $ 0.1131$ &  $ 0.1365$ &   $0.1380$ &  $ 0.1504$  &  $0.0987$
 \\
\lasthline
\end{tabularx}
\caption{  Number of neurons in each recording session that respond differently in salient and non salient image regions, as a fraction of total neurons in the session.  $H_0$ is the null hypothesis in Kolmogorov-Smirnov test. Significance level is $p=0.05$.}
\label{tab:V1ks}
\end{center}
\end{table}
\endgroup

\begingroup
\hyphenpenalty 10000
\exhyphenpenalty 10000
\begin{table}
\footnotesize
\begin{center}
\begin{tabularx}{0.99\linewidth}{L  L L L L L L L}
\midrule[1.5pt]
$H_0$	rejected &    $0.0996$  &  $0.0902$ &   $0.1069$ &   $0.1364$  &  $0.1369$ &   $0.1490$  &  $0.0970$
 \\
\lasthline
\end{tabularx}
\caption{ Number of neurons in each recording session that respond differently in salient and non salient image regions, as a fraction of total neurons in the session.  $H_0$ is the null hypothesis in Kolmogorov-Smirnov test. Significance level is $p=0.05$. Cases where $p$ estimation was inaccurate are rejected.}
\label{tab:V1ksaccu}
\end{center}
\end{table}
\endgroup

\begingroup
\hyphenpenalty 10000
\exhyphenpenalty 10000
\begin{table}
\footnotesize
\begin{center}
\begin{tabularx}{0.99\linewidth}{L  L L L L L L L}
\midrule[1.5pt]
 fail cases & $349$ &  $604$ &  $338$  &  $11$  &  $76$ &  $212$  & $147$
 \\
\lasthline
\end{tabularx}
\caption{When a neuron does not observe the maximum saliency image region in any of the test images, the proposed Kolmogorov-Smirnov test cannot be conducted. We give the number of fail cases in each recording session.}
\label{tab:V1ksfcase}
\end{center}
\end{table}
\endgroup

\begingroup
\hyphenpenalty 10000
\exhyphenpenalty 10000
\begin{table}
\footnotesize
\begin{center}
\begin{tabularx}{0.99\linewidth}{L  L L L L L L L}
\midrule[1.5pt]
$H_0$ rejected &  $0.1071$  &  $0.1028$  &  $0.1173$  &  $0.1367$  &  $0.1392$  &  $0.1536$   & $0.1001$
 \\
\lasthline
\end{tabularx}
\caption{ Number of neurons in each recording session that respond differently in salient and non salient image regions, as a fraction of total cases examined (null hypothesis either accepted or rejected).  $H_0$ is the null hypothesis in Kolmogorov-Smirnov test. Significance level is $p=0.05$.}
\label{tab:V1ksexamined}
\end{center}
\end{table}
\endgroup

\begingroup
\hyphenpenalty 10000
\exhyphenpenalty 10000
\begin{table}
\footnotesize
\begin{center}
\begin{tabularx}{0.99\linewidth}{L  L L L L L L L}
\midrule[1.5pt]
$H_0$ rejected &     $0.1027$  &  $0.0943$  &  $0.1109$  &  $0.1366$  &  $0.1382$  &  $0.1522$ &   $0.0984$
 \\
\lasthline
\end{tabularx}
\caption{ Number of neurons in each recording session that respond differently in salient and non salient image regions, as a fraction of  total cases examined (null hypothesis either accepted or rejected). $H_0$ is the null hypothesis in Kolmogorov-Smirnov test. Significance level is $p=0.05$. Cases where $p$ estimation was inaccurate are rejected.}
\label{tab:V1ksexaminedaccu}
\end{center}
\end{table}
\endgroup

\begingroup
\hyphenpenalty 10000
\exhyphenpenalty 10000
\begin{table}
\footnotesize
\begin{center}
\begin{tabularx}{0.99\linewidth}{L  L L L}
\midrule[1.5pt]
   & Pearson $r$ & Spearman $\rho$ & Kendall $\tau$ \\ \midrule[1.5pt]
5\% percentile & $ -0.0544$ & $-0.0601$ & $-0.0401$ \\ \midrule[1.5pt]
95\% percentile & $0.0713$ & $0.0751$  & $     0.0501$ \\
\lasthline
\end{tabularx}
\caption{ Range of values for correlation coefficients. Data aggregated over all seven recording sessions we analyze in this work.}
\label{tab:V1cc}
\end{center}
\end{table}
\endgroup

\subsection{Retina response shows no signs of visual attention modulation}
\begin{figure}[t]
\setlength\partopsep{-\topsep}
    \centering
    \includegraphics[height=0.4\textheight]{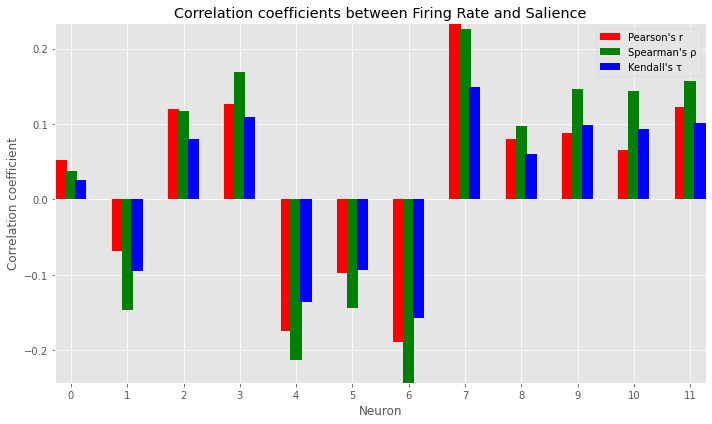}
    \caption{Correlation coefficients between RGC firing rate and visual saliency. Data for twelve CNN modeled RGCs, image stimuli from Toronto set.
}
    \label{fig:Retccs}
\end{figure}

In Figure~\ref{fig:Retccs} we show correlation coefficients $r,\rho,\tau$ for each of the twelve RGCs. We notice that RGCs response is not correlated with visual saliency- coefficients take values near zero (absolute value $<0.25$). Some RGCs get inhibited by salient image regions (negative coefficients).

\begin{figure}[t]
\setlength\partopsep{-\topsep}
    \centering
    \includegraphics[height=0.3\textheight]{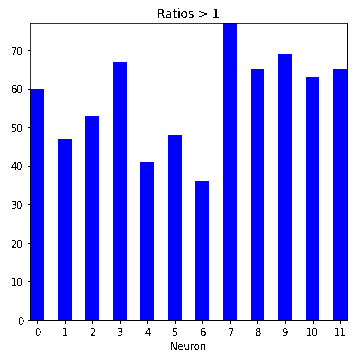}
    \caption{Comparison of baseline response to response in salient image regions: Number of images (cases) in which each RGC showed increased response rate ($r_{img}>1$). Data for all twelve RGCs, image stimuli from Toronto set. For each RGC we examined a total of $120$ cases.
}
    \label{fig:RetRat1}
\end{figure}

\begin{figure}[t]
\setlength\partopsep{-\topsep}
    \centering
    \includegraphics[height=0.3\textheight]{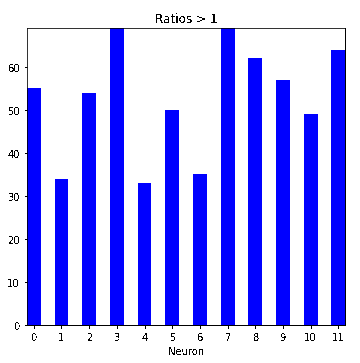}
    \caption{Comparison of baseline response to response in salient image regions: Number of images (cases) in which each RGC showed increased response rate ($r_{rgc}>1$). Data for all twelve RGCs, image stimuli from Toronto set. For each RGC we examined a total of $120$ cases.
}
    \label{fig:RetRat2}
\end{figure}

We turn, in Figures~\ref{fig:RetRat1}, \ref{fig:RetRat2}, to analyzing retina response by the ratios $r_{img},r_{rgc}$. We give the number of ratios with value $>1$, which are the cases where the most salient image region excites the RGC and yields an increase in the FR as compared to the mean RGC FR. These results do not show a particular pattern in RGC response to salient image regions: salient image regions may lead to increase or decrease in the firing rate. All of the examined RGCs yielded a $r>1$ in less than $65\%$ of the images. For both $r_{img},r_{rgc}$, we had values $>1$ in approximately $45\%$ of the images. 

\section{Discussion}
In this work we investigate the emergence of visual attention information in two of the visual system processing stages: Retina and V1. 

We track visual attention information in V1-a fraction of V1 neurons respond differently to salient image regions.  In contrast, analysis of retina response provided no evidence of responsiveness to visually salient regions. 

We follow a computational approach: visual attention is quantified in saliency maps, V1 activity is measured in a large electrophysiology set at single neuron precision. Retina (RGC) response is generated by a deep neural network trained on biological retina response data.

Our findings show that V1 is influenced by saliency. Here we analyze the  spiking response of neurons to natural images- in the case of cortex we take the ground truth neural response,  in the case of retina we simulate the response using a deep model trained using retinal response to natural images. Neuronal response is given at single-neuron precision. Previous studies have favored the cortical versus the retinal formation of saliency maps in living organisms. However, previous studies were based on simple rather than complex stimuli~\cite{treue1999feature}, or analyzed the concerted activity of populations of neurons like imaging studies~\cite{bressler2008top, egner2008neural, gazzaley2012top, giesbrecht2003neural}  or studies using multiunit activity recordings~\cite{bichot2015source}.

This work aims to make our understanding of the biological basis of visual attention more complete, enabling improved visual prostheses. Depending on the implantation site, the relevance of visual attention information changes. Cortical implants may incorporate visual attention to modulate the stimulation of visual attention sensitive cells. In cases were the cells activated by the prosthesis are not directly sensitive to visual attention cues, as in retinal prostheses, assistive systems may be used to direct the attention of implantees to visually salient areas or to enhance salient objects in order to improve their visual presentation to implantees and improve implantees autonomy and visual understanding.

It is important to consider the visually impaired and respond to their needs and habits, to reinforce the acceptability of any intervention. Up until recently, visually impaired people have relied on non-visual cues such as tactile information, canes, guide dogs, communication with their companions etc. In the future, visual prostheses should provide further assistance to the visually impaired. It is consequently important to consider how these interventions can be improved by assistive systems that will process the environment and provide cues to the blind. Such systems are expected to not only boost performance but also improve user acceptability and acquaintance with the interventions, which are all significant for the success of a visual prosthesis. In this direction, visual prosthesis approaches that enhance salient image regions or provide text-reading assistance have been commercialized~\cite{greenberg2017saliency, lauritzen2017text}.

In future work, stronger conclusions may be drawn by improving certain study aspects. Visual attention may differ between species. We analyze mouse response (retinal and cortical) but use a visual attention model tuned for human visual attention. To increase the consistency either a mouse visual attention model may be developed or data from primates, e.g. macaques may be analyzed. Retina Dataset is limited; recording from more RGCs and getting the ground-truth RGC response on Imagenet images will improve retinal data and allow us to use the same statistical methods we applied on the larger V1 set. Finally, V1 conclusions can be augmented by analyzing the cells that respond to visual saliency: Do they form one or more neuronal subtypes? Do these neurons have special and distinct cellular properties e.g. morphology, gene expression?

\bibliographystyle{apalike} 
\bibliography{references} 

\begin{thebibliography}{}

\bibitem[Alevizaki et~al., 2019]{alevizaki2019predicting}
Alevizaki, A., Melanitis, N., and Nikita, K. (2019).
\newblock Predicting eye fixations using computer vision techniques.
\newblock In {\em 2019 IEEE 19th International Conference on Bioinformatics and
  Bioengineering (BIBE)}, pages 309--315. IEEE.

\bibitem[Benetatos et~al., 2021]{benetatos2021assessing}
Benetatos, A., Melanitis, N., and Nikita, K.~S. (2021).
\newblock Assessing vision quality in retinal prosthesis implantees through
  deep learning: Current progress and improvements by optimizing hardware
  design parameters and rehabilitation.
\newblock In {\em 2021 43rd Annual International Conference of the IEEE
  Engineering in Medicine \& Biology Society (EMBC)}, pages 6130--6133. IEEE.

\bibitem[Bichot et~al., 2015]{bichot2015source}
Bichot, N.~P., Heard, M.~T., DeGennaro, E.~M., and Desimone, R. (2015).
\newblock A source for feature-based attention in the prefrontal cortex.
\newblock {\em Neuron}, 88(4):832--844.

\bibitem[Borji, 2018]{borji2018saliency}
Borji, A. (2018).
\newblock Saliency prediction in the deep learning era: Successes, limitations,
  and future challenges.
\newblock {\em arXiv preprint arXiv:1810.03716}.

\bibitem[Bressler et~al., 2008]{bressler2008top}
Bressler, S.~L., Tang, W., Sylvester, C.~M., Shulman, G.~L., and Corbetta, M.
  (2008).
\newblock Top-down control of human visual cortex by frontal and parietal
  cortex in anticipatory visual spatial attention.
\newblock {\em Journal of Neuroscience}, 28(40):10056--10061.

\bibitem[Bruce and Tsotsos, 2007]{bruce2007attention}
Bruce, N. and Tsotsos, J. (2007).
\newblock Attention based on information maximization.
\newblock {\em Journal of Vision}, 7(9):950--950.

\bibitem[Chichilnisky, 2001]{chichilnisky2001simple}
Chichilnisky, E. (2001).
\newblock A simple white noise analysis of neuronal light responses.
\newblock {\em Network: computation in neural systems}, 12(2):199.

\bibitem[Cornia et~al., 2018]{cornia2018predicting}
Cornia, M., Baraldi, L., Serra, G., and Cucchiara, R. (2018).
\newblock Predicting human eye fixations via an lstm-based saliency attentive
  model.
\newblock {\em IEEE Transactions on Image Processing}, 27(10):5142--5154.

\bibitem[da~Cruz et~al., 2016]{da2016five}
da~Cruz, L., Dorn, J.~D., Humayun, M.~S., Dagnelie, G., Handa, J., Barale,
  P.-O., Sahel, J.-A., Stanga, P.~E., Hafezi, F., Safran, A.~B., et~al. (2016).
\newblock Five-year safety and performance results from the argus ii retinal
  prosthesis system clinical trial.
\newblock {\em Ophthalmology}, 123(10):2248--2254.

\bibitem[Deng et~al., 2009]{deng2009imagenet}
Deng, J., Dong, W., Socher, R., Li, L.-J., Li, K., and Fei-Fei, L. (2009).
\newblock Imagenet: A large-scale hierarchical image database.
\newblock In {\em 2009 IEEE conference on computer vision and pattern
  recognition}, pages 248--255. Ieee.

\bibitem[Egner et~al., 2008]{egner2008neural}
Egner, T., Monti, J.~M., Trittschuh, E.~H., Wieneke, C.~A., Hirsch, J., and
  Mesulam, M.-M. (2008).
\newblock Neural integration of top-down spatial and feature-based information
  in visual search.
\newblock {\em Journal of Neuroscience}, 28(24):6141--6151.

\bibitem[Fernandez, 2018]{fernandez2018development}
Fernandez, E. (2018).
\newblock Development of visual neuroprostheses: trends and challenges.
\newblock {\em Bioelectronic medicine}, 4(1):1--8.

\bibitem[Fern{\'a}ndez et~al., 2021]{fernandez2021visual}
Fern{\'a}ndez, E., Alfaro, A., Soto-S{\'a}nchez, C., Gonzalez-Lopez, P.,
  Lozano, A.~M., Pe{\~n}a, S., Grima, M.~D., Rodil, A., G{\'o}mez, B., Chen,
  X., et~al. (2021).
\newblock Visual percepts evoked with an intracortical 96-channel
  microelectrode array inserted in human occipital cortex.
\newblock {\em The Journal of clinical investigation}, 131(23).

\bibitem[Gan et~al., 2015]{gan2015devnet}
Gan, C., Wang, N., Yang, Y., Yeung, D.-Y., and Hauptmann, A.~G. (2015).
\newblock Devnet: A deep event network for multimedia event detection and
  evidence recounting.
\newblock In {\em Proceedings of the IEEE Conference on Computer Vision and
  Pattern Recognition}, pages 2568--2577.

\bibitem[Gazzaley and Nobre, 2012]{gazzaley2012top}
Gazzaley, A. and Nobre, A.~C. (2012).
\newblock Top-down modulation: bridging selective attention and working memory.
\newblock {\em Trends in cognitive sciences}, 16(2):129--135.

\bibitem[Giesbrecht et~al., 2003]{giesbrecht2003neural}
Giesbrecht, B., Woldorff, M.~G., Song, A.~W., and Mangun, G.~R. (2003).
\newblock Neural mechanisms of top-down control during spatial and feature
  attention.
\newblock {\em Neuroimage}, 19(3):496--512.

\bibitem[Gottlieb et~al., 1998]{gottlieb1998representation}
Gottlieb, J.~P., Kusunoki, M., and Goldberg, M.~E. (1998).
\newblock The representation of visual salience in monkey parietal cortex.
\newblock {\em Nature}, 391(6666):481--484.

\bibitem[Greenberg et~al., 2017]{greenberg2017saliency}
Greenberg, R., Horsager, A., Humayun, M.~S., McClure, K.~H., McMahon, M.~J.,
  Meilstrup, P., Parikh, N., Roy, A., Weiland, J.~D., Zhou, C., et~al. (2017).
\newblock Saliency-based apparatus and methods for visual prostheses.
\newblock US Patent 9,795,786.

\bibitem[Gupta et~al., 2020]{gupta2020salient}
Gupta, A.~K., Seal, A., Prasad, M., and Khanna, P. (2020).
\newblock Salient object detection techniques in computer vision—a survey.
\newblock {\em Entropy}, 22(10):1174.

\bibitem[He et~al., 2016]{he2016deep}
He, K., Zhang, X., Ren, S., and Sun, J. (2016).
\newblock Deep residual learning for image recognition.
\newblock In {\em Proceedings of the IEEE conference on computer vision and
  pattern recognition}, pages 770--778.

\bibitem[Huang et~al., 2015]{huang2015salicon}
Huang, X., Shen, C., Boix, X., and Zhao, Q. (2015).
\newblock Salicon: Reducing the semantic gap in saliency prediction by adapting
  deep neural networks.
\newblock In {\em Proceedings of the IEEE international conference on computer
  vision}, pages 262--270.

\bibitem[Itti et~al., 1998]{itti1998model}
Itti, L., Koch, C., and Niebur, E. (1998).
\newblock A model of saliency-based visual attention for rapid scene analysis.
\newblock {\em IEEE Transactions on pattern analysis and machine intelligence},
  20(11):1254--1259.

\bibitem[Lauritzen et~al., 2017]{lauritzen2017text}
Lauritzen, T., Dorn, J.~D., Greenberg, R.~J., Harris, J., and Sahel, J.~A.
  (2017).
\newblock Text reading and translation in a visual prosthesis.
\newblock US Patent 9,715,837.

\bibitem[Li and Yu, 2015]{li2015visual}
Li, G. and Yu, Y. (2015).
\newblock Visual saliency based on multiscale deep features.
\newblock In {\em Proceedings of the IEEE conference on computer vision and
  pattern recognition}, pages 5455--5463.

\bibitem[Li, 2002]{li2002saliency}
Li, Z. (2002).
\newblock A saliency map in primary visual cortex.
\newblock {\em Trends in cognitive sciences}, 6(1):9--16.

\bibitem[Lozano et~al., 2018]{lozano20183d}
Lozano, A., Soto-Sanchez, C., Garrigos, J., Mart{\'\i}nez, J.~J.,
  Ferr{\'a}ndez, J.~M., and Fernandez, E. (2018).
\newblock A 3d convolutional neural network to model retinal ganglion cell’s
  responses to light patterns in mice.
\newblock {\em International journal of neural systems}, 28(10):1850043.

\bibitem[Massey~Jr, 1951]{massey1951kolmogorov}
Massey~Jr, F.~J. (1951).
\newblock The kolmogorov-smirnov test for goodness of fit.
\newblock {\em Journal of the American statistical Association},
  46(253):68--78.

\bibitem[McIntosh et~al., 2016]{mcintosh2016deep}
McIntosh, L., Maheswaranathan, N., Nayebi, A., Ganguli, S., and Baccus, S.
  (2016).
\newblock Deep learning models of the retinal response to natural scenes.
\newblock {\em Advances in neural information processing systems}, 29.

\bibitem[Melanitis et~al., 2021]{melanitis2021using}
Melanitis, N., Nakopoulos, G., Lozano, A., Soto-Sanchez, C., Fernandez, E., and
  Nikita, K.~S. (2021).
\newblock Using biologically-inspired image features to model retinal response:
  Evidence from biological datasets.
\newblock In {\em 2021 43rd Annual International Conference of the IEEE
  Engineering in Medicine \& Biology Society (EMBC)}, pages 3378--3381. IEEE.

\bibitem[Melanitis and Nikita, 2019]{melanitis2019biologically}
Melanitis, N. and Nikita, K.~S. (2019).
\newblock Biologically-inspired image processing in computational retina
  models.
\newblock {\em Computers in biology and medicine}, 113:103399.

\bibitem[Moore and Zirnsak, 2015]{moore2015and}
Moore, T. and Zirnsak, M. (2015).
\newblock The what and where of visual attention.
\newblock {\em Neuron}, 88(4):626--628.

\bibitem[Papadopoulos et~al., 2021]{papadopoulos2021machine}
Papadopoulos, N., Melanitis, N., Lozano, A., Soto-Sanchez, C., Fernandez, E.,
  and Nikita, K.~S. (2021).
\newblock Machine learning method for functional assessment of retinal models.
\newblock In {\em 2021 43rd Annual International Conference of the IEEE
  Engineering in Medicine \& Biology Society (EMBC)}, pages 4293--4296. IEEE.

\bibitem[Pouratian et~al., 2019]{pouratian2019early}
Pouratian, N., Yoshor, D., Niketeghad, S., Dornm, J., and Greenberg, R. (2019).
\newblock Early feasibility study of a neurostimulator to create artificial
  vision.
\newblock {\em Neurosurgery}, 66(Supplement\_1):nyz310\_146.

\bibitem[Press et~al., 2007]{press2007numerical}
Press, W.~H., Teukolsky, S.~A., Vetterling, W.~T., and Flannery, B.~P. (2007).
\newblock {\em Numerical recipes 3rd edition: The art of scientific computing}.
\newblock Cambridge university press.

\bibitem[Robinson and Petersen, 1992]{robinson1992pulvinar}
Robinson, D.~L. and Petersen, S.~E. (1992).
\newblock The pulvinar and visual salience.
\newblock {\em Trends in neurosciences}, 15(4):127--132.

\bibitem[Rother et~al., 2006]{rother2006autocollage}
Rother, C., Bordeaux, L., Hamadi, Y., and Blake, A. (2006).
\newblock Autocollage.
\newblock {\em ACM transactions on graphics (TOG)}, 25(3):847--852.

\bibitem[Rutishauser et~al., 2004]{rutishauser2004bottom}
Rutishauser, U., Walther, D., Koch, C., and Perona, P. (2004).
\newblock Is bottom-up attention useful for object recognition?
\newblock In {\em Proceedings of the 2004 IEEE Computer Society Conference on
  Computer Vision and Pattern Recognition, 2004. CVPR 2004.}, volume~2, pages
  II--II. IEEE.

\bibitem[Shimoda and Yanai, 2016]{shimoda2016distinct}
Shimoda, W. and Yanai, K. (2016).
\newblock Distinct class-specific saliency maps for weakly supervised semantic
  segmentation.
\newblock In {\em European conference on computer vision}, pages 218--234.
  Springer.

\bibitem[Simakov et~al., 2008]{simakov2008summarizing}
Simakov, D., Caspi, Y., Shechtman, E., and Irani, M. (2008).
\newblock Summarizing visual data using bidirectional similarity.
\newblock In {\em 2008 IEEE Conference on Computer Vision and Pattern
  Recognition}, pages 1--8. IEEE.

\bibitem[Smyth et~al., 2003]{smyth2003receptive}
Smyth, D., Willmore, B., Baker, G.~E., Thompson, I.~D., and Tolhurst, D.~J.
  (2003).
\newblock The receptive-field organization of simple cells in primary visual
  cortex of ferrets under natural scene stimulation.
\newblock {\em Journal of Neuroscience}, 23(11):4746--4759.

\bibitem[Sousa et~al., 2009]{sousa2009bioelectronic}
Sousa, L.~A. et~al. (2009).
\newblock {\em Bioelectronic vision: retina models, evaluation metrics, and
  system design}, volume~3.
\newblock World Scientific.

\bibitem[Stringer et~al., 2019]{stringer2019high}
Stringer, C., Pachitariu, M., Steinmetz, N., Carandini, M., and Harris, K.~D.
  (2019).
\newblock High-dimensional geometry of population responses in visual cortex.
\newblock {\em Nature}, 571(7765):361--365.

\bibitem[Touryan et~al., 2005]{touryan2005spatial}
Touryan, J., Felsen, G., and Dan, Y. (2005).
\newblock Spatial structure of complex cell receptive fields measured with
  natural images.
\newblock {\em Neuron}, 45(5):781--791.

\bibitem[Treue, 2001]{treue2001neural}
Treue, S. (2001).
\newblock Neural correlates of attention in primate visual cortex.
\newblock {\em Trends in neurosciences}, 24(5):295--300.

\bibitem[Treue, 2003]{treue2003visual}
Treue, S. (2003).
\newblock Visual attention: the where, what, how and why of saliency.
\newblock {\em Current opinion in neurobiology}, 13(4):428--432.

\bibitem[Treue and Trujillo, 1999]{treue1999feature}
Treue, S. and Trujillo, J. C.~M. (1999).
\newblock Feature-based attention influences motion processing gain in macaque
  visual cortex.
\newblock {\em Nature}, 399(6736):575--579.

\bibitem[Weiland and Humayun, 2014]{weiland2014retinal}
Weiland, J.~D. and Humayun, M.~S. (2014).
\newblock Retinal prosthesis.
\newblock {\em IEEE Transactions on Biomedical Engineering}, 61(5):1412--1424.

\bibitem[Wu et~al., 2013]{wu2013scale}
Wu, R., Yu, Y., and Wang, W. (2013).
\newblock Scale: Supervised and cascaded laplacian eigenmaps for visual object
  recognition based on nearest neighbors.
\newblock In {\em Proceedings of the IEEE conference on computer vision and
  pattern recognition}, pages 867--874.

\end{thebibliography}
\end{document}